\documentclass[aps,prd,preprint,twoside,tightenlines,showpacs,amsmath,amssymb,nofootinbib,floatfix,superscriptaddress]{revtex4}
\usepackage{graphicx}
\usepackage{amsfonts}
\usepackage{color}
\usepackage[colorlinks=true,urlcolor=blue,linkcolor=blue,citecolor=blue]{hyperref}
\usepackage[T1]{fontenc}
\usepackage{lmodern}

\begin{document}
\renewcommand{\theenumi}{(\alph{enumi})}

\title{\boldmath $B^0$ and $B^0_s$ decays into $J/\psi$ plus a scalar or vector meson}
\date{\today}
\author{M. ~Bayar}
\affiliation{Department of Physics, Kocaeli University, 41380, Izmit, Turkey}
\affiliation{Departamento de
F\'{\i}sica Te\'orica and IFIC, Centro Mixto Universidad de
Valencia-CSIC Institutos de Investigaci\'on de Paterna, Aptdo.
22085, 46071 Valencia, Spain}
\author{W.H. Liang}
\email{liangwh@gxnu.edu.cn}
\affiliation{Department of Physics, Guangxi Normal University, Guilin 541004, China}
\affiliation{Departamento de
F\'{\i}sica Te\'orica and IFIC, Centro Mixto Universidad de
Valencia-CSIC Institutos de Investigaci\'on de Paterna, Aptdo.
22085, 46071 Valencia, Spain}
\affiliation{Kavli Institute for Theoretical Physics China, CAS, Beijing 100190, China}
\author{E.~Oset}
\affiliation{Departamento de
F\'{\i}sica Te\'orica and IFIC, Centro Mixto Universidad de
Valencia-CSIC Institutos de Investigaci\'on de Paterna, Aptdo.
22085, 46071 Valencia, Spain}
\affiliation{Institute of Modern Physics, Chinese Academy of Sciences, Lanzhou
730000, China}

\begin{abstract}
We extend a recent approach to describe the $B^0$ and $B^0_s$ decays into $J/\psi$ $f_0(500)$ and $J/\psi$ $f_0(980)$, relating it to the  $B^0$ and $B^0_s$ decays into $J/\psi$ and a vector meson, $\phi$, $\rho$, $K^*$. In addition the $B^0$ and $B^0_s$ decays into $J/\psi$ and $\kappa(800)$ are evaluated and compared to the $K^*$ vector production. The rates obtained are in  agreement with available experiment while predictions are made for the $J/\psi$ plus $\kappa(800)$ decay.
\end{abstract}

\maketitle

\section{Introduction}

The $B^0_s$ decays into $J/\psi$ plus $f_0(500)$ or $f_0(980)$ are capturing the attention of both experiment and theory. A striking result observed in LHCb is that in the $B^0_s$ decay a pronounced peak for the $f_0(980)$ is observed \cite{Aaij:2011fx} while no appreciable signal is seen for the $f_0(500)$. These results have been corroborated by the Belle \cite{Li:2011pg},
CDF \cite{Aaltonen:2011nk}, D0 \cite{Abazov:2011hv}, and again LHCb  \cite{LHCb:2012ae,Aaij:2014emv} collaborations. Conversely, in \cite{Aaij:2013zpt}, the $\bar B^0$ into $J/\psi$ and $\pi^+ \pi^-$  is investigated, and a signal is seen for the $f_0(500)$ production while only a very small fraction is observed for the $f_0(980)$ production.

 Estimations of the order of magnitude of rates for some of these reactions have been done using light cone QCD sum rules under the factorization assumption \cite{Colangelo:2010bg}. Also, the experimental data have served as a basis of discussion on the possible nature of the scalar mesons as a $q \bar q$ or tetraquark \cite{Stone:2013eaa}.

More recently, a simple approach based on the final state interaction of mesons provided by the chiral unitary approach has been applied that allows us to calculate all these rates relative to one of them \cite{weihong}. The work isolates the dominant weak decay mechanism into $J/\psi$ and a $q \bar q$ pair. After this, the $q \bar q$ pair is hadronized, and meson-meson pairs are produced with a certain weight. These mesons are then allowed to interact, and for this, the chiral unitary approach for meson-meson interaction \cite{npa,ramonet,kaiser,markushin,juanito,rios} is used. This approach uses a full unitary scheme by means of the Bethe-Salpeter equation in coupled channels \cite{Kaiser:1995eg,Oller:2000ma}, extracting the kernel, or potential, from the chiral Lagrangians \cite{Gasser:1983yg,Bernard:1995dp}. The success of this approach to deal with meson-meson interaction and with reactions in which the $f_0(500)$ or $f_0(980)$ and other resonances are produced is remarkable (see \cite{weihong} for a detailed list of reactions studied), but the closest ones are the $J/\psi \to \phi (\omega) \pi \pi$ where different signals for the $f_0(500)$, $f_0(980)$ are observed depending on the reaction \cite{Meissner:2000bc,chiangpalo,Lahde:2006wr,Hanhart:2007bd,Roca:2012cv}. The idea of these latter works in which the final state interaction of pairs of mesons is explicitly taken into account has been also followed in weak decays similar to those discussed above, concretely in  the $B \to \pi \pi K $ decay \cite{robert,bruno}.

   The related experimental work on vector meson production is more abundant. The $B^0_s \rightarrow J/\psi \bar{K}^{*0}$ is studied recently in \cite{Aaltonen:2011nk,Aaij:2012nh} [see the particle data book (PDG) for more experiments \cite{pdg}], and the $B^0 \rightarrow J/\psi {K}^{*0}$ is studied in \cite{Aubert:2007qea,Aubert:2004rz} among others \cite{pdg}. The $\rho$ production is studied as a part of the spectra of the $B^0$ decay into $J/\psi$ and $\pi^+ \pi^-$ in \cite{Aaij:2013zpt,Aaij:2014siy}.

   In the present paper we review shortly the work of \cite{weihong} and complement it by evaluating the rates of $B^0$ and $B^0_s$ decays into $J/\psi$ and a vector, $\phi$, $\rho$, $K^*$. In addition we also evaluate the rates of $B^0$ and $B^0_s$ decays into $J/\psi$ and $\kappa$. This allows us to compare the rates for $K^*$ and $\kappa$ production, as well as $f_0(500)$, $f_0(980)$ with $\rho$ production, and $\kappa$ production with $f_0(500)$, $f_0(980)$ production. The work exploits flavor symmetries and dynamics of meson-meson interaction and  factorizes the matrix elements of the weak process, which are not explicitly evaluated. These latter ones are shared by different reactions such that at the end, by using only two rates from experiment, we can produce all the mass distributions for all the different reactions possible.

 \section{Formalism for scalar meson production}

\begin{figure}[t!]\centering
\includegraphics[height=3.0cm,keepaspectratio]{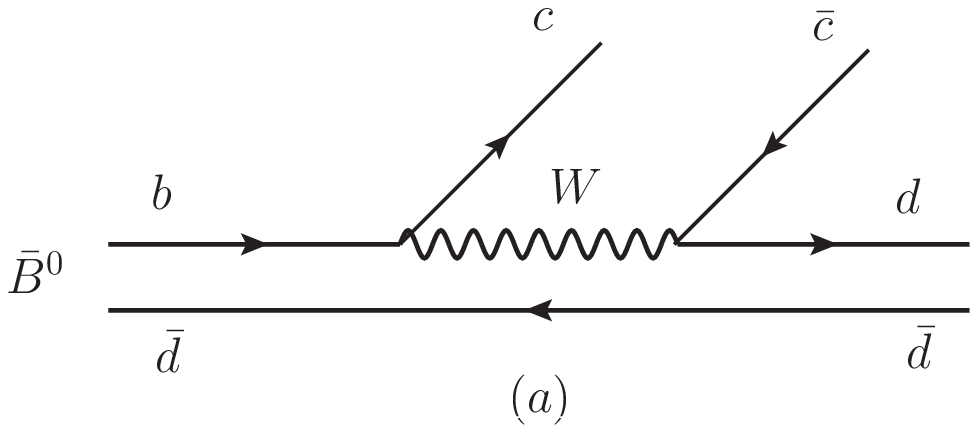}
\includegraphics[height=3.2cm,keepaspectratio]{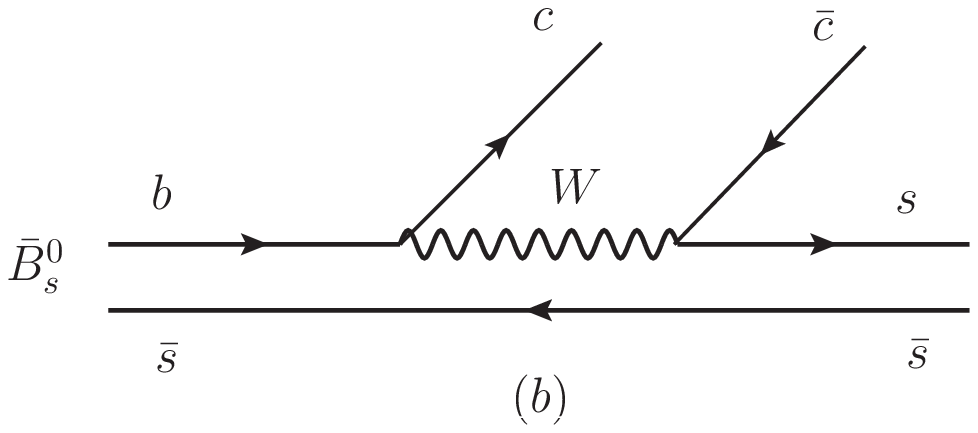}
\includegraphics[height=3.2cm,keepaspectratio]{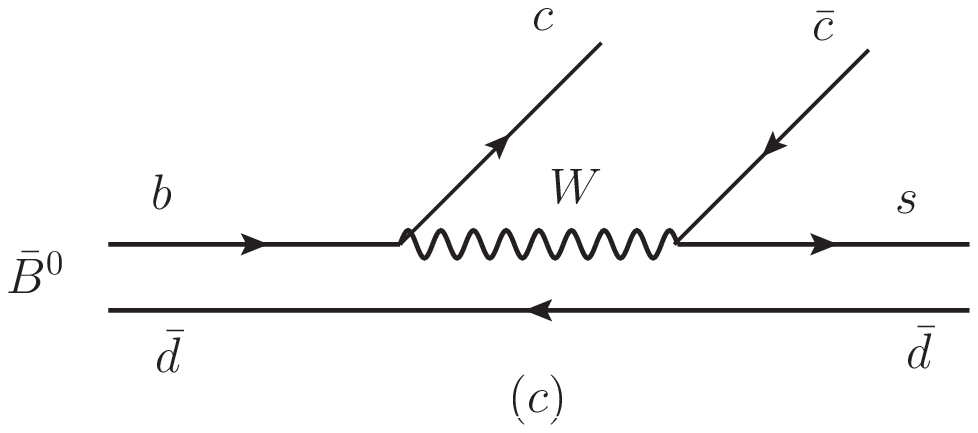}
\includegraphics[height=3.2cm,keepaspectratio]{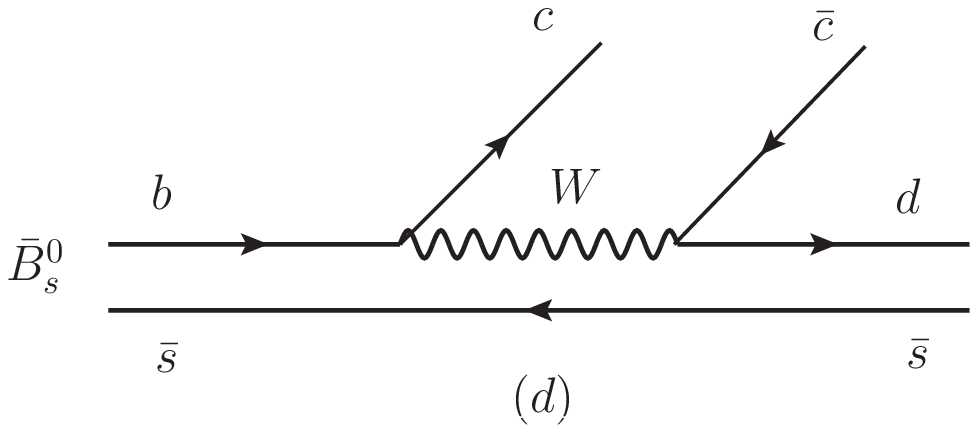}
\caption{Diagrams for the decay of $\bar B^0$ and $\bar B^0_s$ into $J/\psi$ and a primary $q\bar q$ pair. (a) Cabbibo suppressed ${\bar B}^0$ decay,  (b) Cabbibo favored
${\bar B}^0_s$ decay, (c) Cabbibo favored ${\bar B}^0$ decay, (d) Cabbibo suppressed ${\bar B}^0_s$ decay.
\label{fig:fig1}}
\end{figure}

\begin{figure}[h!]\centering
\includegraphics[height=1.85cm,keepaspectratio]{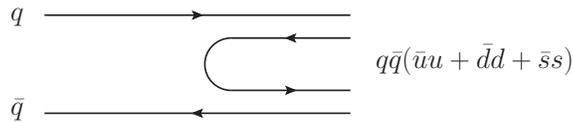}
\caption{Schematic representation of the hadronization $q\bar q \to q\bar q (\bar u u +\bar d d +\bar s s)$.\label{fig:fig2}}
\end{figure}

Following \cite{Aaij:2011fx,Li:2011pg,Aaltonen:2011nk,Abazov:2011hv,LHCb:2012ae,Aaij:2013zpt,
Aaij:2014emv,Stone:2013eaa,Aaij:2014siy}, we take the dominant weak mechanism for $\bar{B}^0$ and $\bar{B}^0_s$ decays as depicted in Fig.~\ref{fig:fig1}. The case of $B^0$ ($B^0_s$) is
identical to that of $\bar B^0$ ($\bar B^0_s$), changing the particles by their antiparticles. In Fig.~\ref{fig:fig1}(a), in addition to the $J/\psi$, a primary pair of $d \bar d$ quarks are produced from the $B^0$ decay, while an $s \bar s$ pair is produced in the case of the $B^0_s$ decay, Fig.~\ref{fig:fig1}(b). These two cases are those studied in \cite{weihong}. In addition, we can also produce a $s \bar d$ pair in the  $\bar{B}^0$ decay and a $d \bar s$ pair in the $\bar{B}^0_s$ decay, Figs.~\ref{fig:fig1}(c) and \ref{fig:fig1}(d). These two latter cases are new and we study them here. If we look for production of scalar mesons, $f_0(500)$,  $f_0(980)$, and $\kappa$, one identifies them by looking at $\pi^+ \pi^-$ production in the case of $f_0(500)$,  $f_0(980)$, and $\pi K$ for the case of the $\kappa$. We have to produce two mesons, which means that the $q \bar q$ pair must hadronize. To accomplish this, we follow the approach of
 \cite{alberzou} and complement the primary $q \bar q$ pair by another $\bar q q$ pair with the quantum numbers of the vacuum $\bar u u +\bar d d +\bar s s$ (see Fig.~\ref{fig:fig2}). Then, we realize that the $q \bar q$ matrix $M$

\begin{equation}\label{eq:1}
M=\left(
           \begin{array}{ccc}
             u\bar u & u \bar d & u\bar s \\
             d\bar u & d\bar d & d\bar s \\
             s\bar u & s\bar d & s\bar s \\
           \end{array}
         \right)
\end{equation}
 has the property
\begin{equation}\label{eq:2}
M\cdot M=M \times (\bar u u +\bar d d +\bar s s).
\end{equation}

Now, in terms of mesons, neglecting the $\eta_1$ singlet that corresponds mostly to the $\eta'$, which we omit in the coupled channels because of its large mass, the matrix $M$ corresponds to
\begin{equation}\label{eq:3}
\phi=\left(
           \begin{array}{ccc}
            \frac{1}{\sqrt{2}}\pi^0 +\frac{1}{\sqrt{6}} \eta & \pi^+  & K^+\\
             \pi^- &  -\frac{1}{\sqrt{2}}\pi^0 +\frac{1}{\sqrt{6}} \eta & K^0 \\
             K^- & \bar K^0 & -\frac{2}{\sqrt{6}} \eta \\
           \end{array}
         \right),
\end{equation}
where $\eta$ is actually $\eta_8$, but can be considered the $\eta$ for practical purposes.   Hence, in terms of two pseudoscalars we have the correspondence
\begin{align}\label{eq:4}
d\bar d (\bar u u +\bar d d +\bar s s) & \equiv  \left( \phi \cdot \phi \right)_{22}=\pi^- \pi^+ +\frac{1}{2}\pi^0 \pi^0
-\frac{1}{\sqrt{3}}\pi^0 \eta +K^0 \bar K^0 +\frac{1}{6}\eta \eta ,\nonumber \\
s\bar s (\bar u u +\bar d d +\bar s s) & \equiv  \left( \phi \cdot \phi \right)_{33}=K^- K^+ + K^0 \bar K^0 +\frac{4}{6}\eta \eta  , \\
s\bar d (\bar u u +\bar d d +\bar s s) & \equiv  \left( \phi \cdot \phi \right)_{32}=K^- \pi^+ -\frac{1}{\sqrt{2}}\bar K^0 \pi^0
-\frac{1}{\sqrt{6}} \eta \bar K^0,\nonumber \\
d\bar s (\bar u u +\bar d d +\bar s s) & \equiv  \left( \phi \cdot \phi \right)_{23}=\pi^- K^+  -\frac{1}{\sqrt{2}} K^0 \pi^0
-\frac{1}{\sqrt{6}} \eta K^0. \nonumber
\end{align}

  The diagrams of Fig. \ref{fig:fig1} share the same dynamics and are only differentiated by the different matrix element of the Cabbibo-Kobayashi-Maskawa (CKM) matrix. From the second $qqW$ vertex in the diagrams, we have $V_{cd}$ in Figs. \ref{fig:fig1}(a) and \ref{fig:fig1}(d) and $V_{cs}$ in Figs. \ref{fig:fig1}(b) and \ref{fig:fig1}(c). These matrix elements are related to the Cabbibo angle
\begin{align}\label{eq:Cabbibo}
&V_{cd}=-\sin\theta_c =-0.22534,\nonumber \\
&V_{cs}=\cos\theta_c =0.97427.
\end{align}

\begin{figure}[t!]\centering
\includegraphics[height=3.0cm,keepaspectratio]{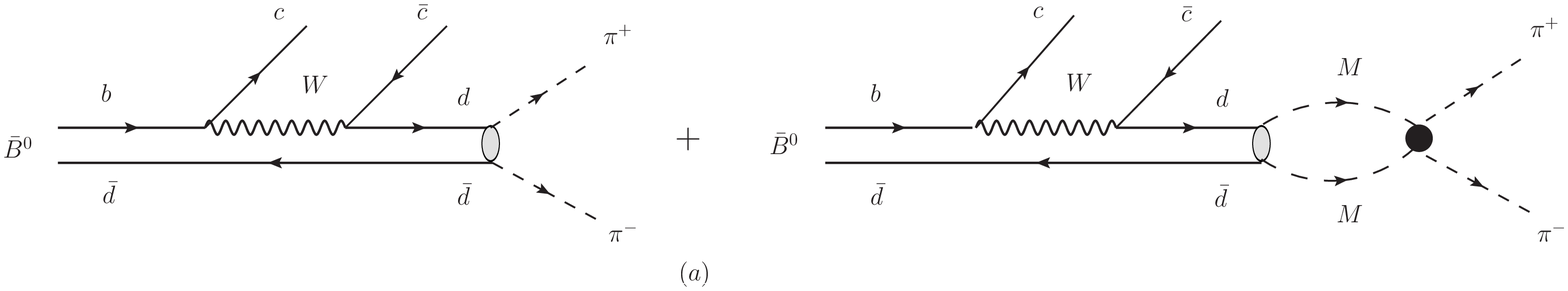}
\includegraphics[height=3.0cm,keepaspectratio]{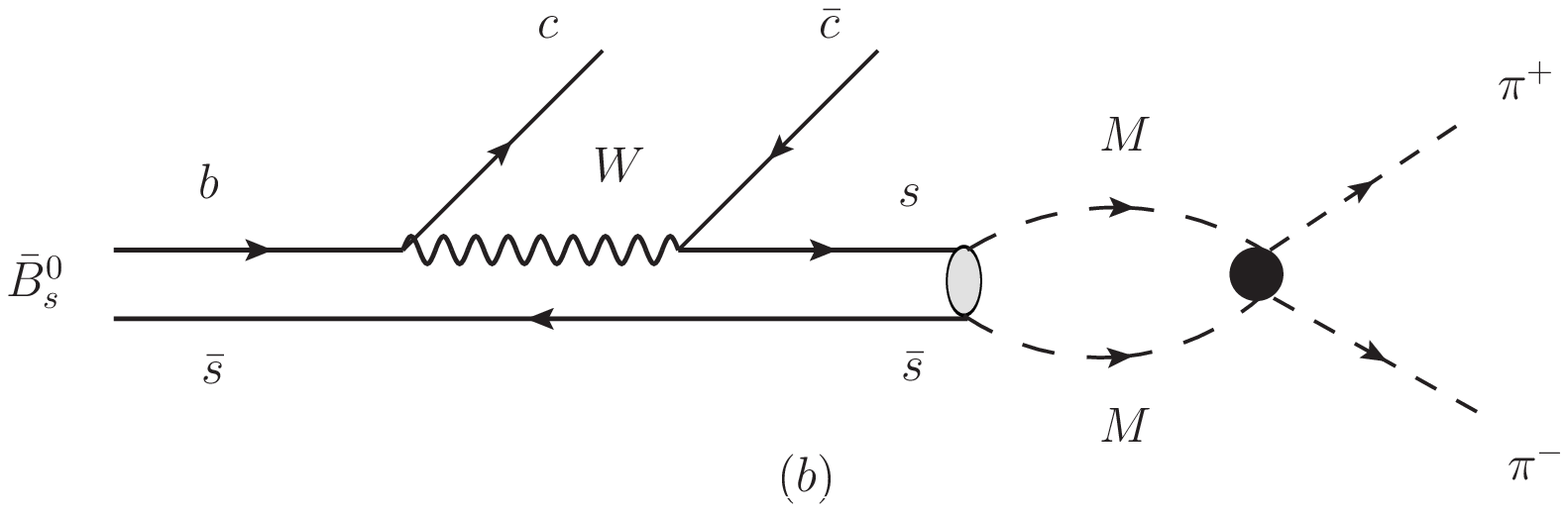}
\includegraphics[height=3.0cm,keepaspectratio]{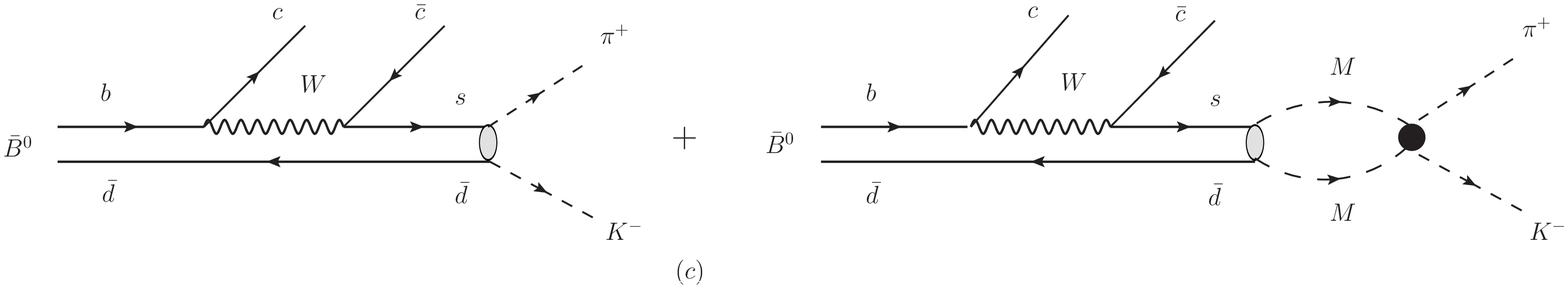}
\includegraphics[height=3.0cm,keepaspectratio]{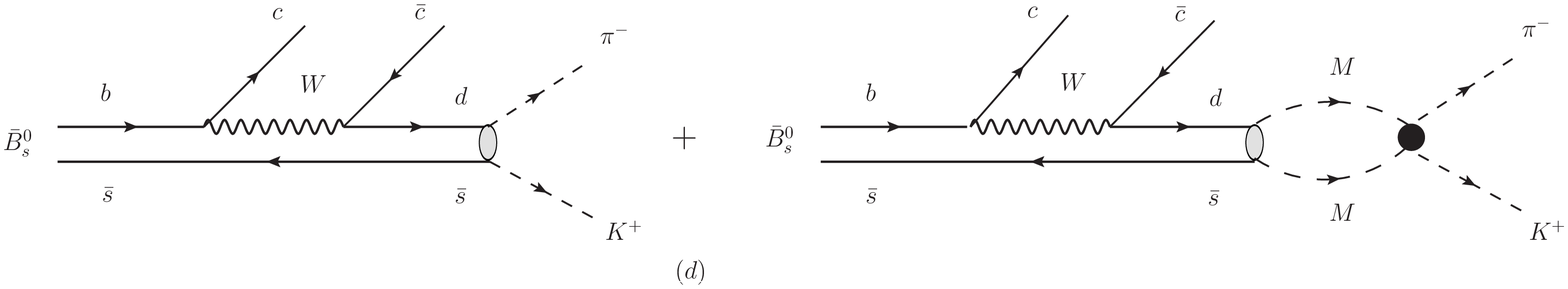}
\caption{Diagrammatic representations of $\pi^+ \pi^-$, $\pi^+ K^-$ and $\pi^- K^+$ via direct plus rescattering mechanisms in $\bar B^0$ and $\bar B ^0_s$ decays. (a),(b),(c) and (d) correspond to the diagrams of Fig.~\ref{fig:fig1}
upon hadronization of $q \bar q$ into mesons and further rescattering. \label{fig:fig3}}
\end{figure}

The next step consists of allowing the pair of mesons originated in the first step to interact among themselves and their coupled channels, since this interaction is what gives rise dynamically to the low-lying scalar mesons in chiral unitary theory. This is depicted diagrammatically in Fig. \ref{fig:fig3} for $\pi^+ \pi^-$ and $\pi^+ K^-$ or $\pi^- K^+$ production.

We can see that $\pi^+ \pi^-$ is obtained in the first step in the $\bar B^0$ decay but not in $\bar B^0_s$ decay. In this latter case, upon rescattering of $K\bar K$ we also can get $\pi^+ \pi^-$ in the final state. Since the $f_0(980)$ couples strongly to $K\bar K$ and the $f_0(500)$ to $\pi \pi$, the meson-meson decomposition of Eq.~(\ref{eq:4}) is already hinting that the $\bar B^0$ decay will be dominated by $f_0(500)$ production and $\bar B^0_s$ decay by $f_0(980)$ production. This is indeed what was found in \cite{weihong}.

The primary production and rescattering of the mesons is taken into account as follows:
Let us call $V_P$ the production vertex  containing all dynamical factors common to the four reactions. The $\pi^+ \pi^-$ or $\pi K$ production will proceed via primary production or final state interaction as depicted in Fig.~\ref{fig:fig3}.

The amplitudes for $\pi^+ \pi^-$ and $\pi K$ production are given by
\begin{align}\label{eq:5}
t(\bar B^0 \to J/\psi \pi^+ \pi^-) & = V_P V_{cd}
(
1+G_{\pi^+ \pi^-} t_{\pi^+ \pi^- \to \pi^+ \pi^-} +\frac{1}{2} \frac{1}{2} G_{\pi^0 \pi^0} t_{\pi^0 \pi^0 \to \pi^+ \pi^-}  \nonumber \\
&~~ +G_{K^0 \bar K^0} t_{K^0 \bar K^0 \to \pi^+ \pi^-} +\frac{1}{6} \frac{1}{2} G_{\eta \eta} t_{\eta \eta \to \pi^+ \pi^-}) ,\nonumber \\
t(\bar B^0_s \to J/\psi \pi^+ \pi^-) & = V_P V_{cs} ( G_{K^+ K^-} t_{K^+ K^- \to \pi^+ \pi^-} +  G_{K^0 \bar K^0} t_{K^0 \bar K^0 \to \pi^+ \pi^-}  +\frac{4}{6} \frac{1}{2} G_{\eta \eta} t_{\eta \eta \to \pi^+ \pi^-} )~, \nonumber \\
t(\bar B^0 \to J/\psi \pi^+ K^-) & = V_P V_{cs}
(
1+G_{\pi^+ K^-} t_{\pi^+ K^- \to \pi^+ K^-} \nonumber \\
&~~ -\frac{1}{\sqrt{2}} G_{\pi^0 \bar K^0} t_{\pi^0 \bar K^0 \to \pi^+ K^-} - \frac{1}{\sqrt{6}} G_{\eta \bar K^0} t_{\eta \bar K^0 \to \pi^+ K^-}), \\
t(\bar B^0_s \to J/\psi \pi^-  K^+) & = V_P V_{cd} (1+ G_{\pi^- K^+} t_{\pi^- K^+ \to \pi^- K^+} \nonumber \\
&~~ -\frac{1}{\sqrt{2}}  G_{\pi^0 K^0} t_{\pi^0 K^0 \to \pi^- K^+}  -\frac{1}{\sqrt{6}} G_{\eta K^0} t_{\eta K^0 \to \pi^- K^+} )~, \nonumber
\end{align}
where the $G_i$'s are the loop functions of two meson propagators
\begin{equation}
G_i (s) = i \int\frac{d^{4}q}{(2\pi)^{4}}\frac{1}{(P-q)^{2}-m^2_1+i\varepsilon}\,\frac{1}{q^{2}-m^2_2+i\varepsilon},
\label{eq:G}
\end{equation}
with $m_1, ~m_2$ the masses of the mesons in the $i$ channel, $q$ the four-momentum of one meson, and $P$ the total four-momentum of the system, thus, $s=P^2$.
The integral is performed integrating exactly the $q^0$ variable and implementing a cutoff $\Lambda$ of the order on $1~ {\rm{GeV}} /c$ for the three-momentum ( in \cite{weihong} it was taken as $600 ~{\rm MeV}/c$ after including explicitly the $\eta \eta$ channel).
The elements $t_{ij}$ are the scattering matrices for transitions of channel $i$ to $j$. According to \cite{npa}, this matrix is given by
\begin{equation}\label{eq:BSeq}
t = [1-VG]^{-1} V,
\end{equation}
and the $V$ matrix is taken from \cite{npa} complemented with the matrix elements of the $\eta \eta$ channels, which we have taken from \cite{danydan}. Explicit forms of the potential are given in \cite{weihong}. Note that we include the factor $1/2$ before the $G$ function in the case of identical particles. As discussed in \cite{weihong}, the unitary normalization of the identical states is used to get the $V$ and $T$ matrices by means of Eq. (\ref{eq:BSeq}), but the good normalization must be used for the $t$ matrices in Eqs. (\ref{eq:5}).

  In addition, to deal with the $\kappa$, we have to solve the Bethe-Salpeter equation of Eq. (\ref{eq:BSeq}) with the channels $\pi^- K^+$, $\pi^0 K^0$, and $\eta K^0$. These channels are numbered orderly as 1, 2, and 3. The matrix elements projecting into the $S$ wave are taken from \cite{danydan} and given by
  \begin{align}\label{eq:Vkernel}
&V_{11}=-\frac{1}{4f^2}s,
~&&V_{12}=-\frac{1}{2\sqrt{2}f^2}(-\frac{3}{2}s+m_{\pi}^2+m_K^2), \nonumber \\
&V_{13}=-\frac{1}{6\sqrt{6}f^2}(-\frac{9}{2}s+9m_K^2-\frac{1}{2}m_{\pi}^2+\frac{3}{2}m_{\eta}^2),
~&&V_{22}=-\frac{1}{12f^2}(-\frac{3}{2}s+3m_K^2+3m_{\pi}^2),  \\
&V_{23}=-\frac{1}{12\sqrt{3}f^2}(\frac{9}{2}s-m_K^2-\frac{7}{2}m_{\pi}^2-\frac{3}{2}m_{\eta}^2),
~&&V_{33}=-\frac{1}{12f^2}(-\frac{9}{2}s+9m_K^2+3m_{\eta}^2-2m_{\pi}^2),\nonumber
\end{align}
with $f=93~{\rm MeV}$ the pion decay constant.

In Eq.~(\ref{eq:5}) we made use of the fact that both the  $f_0(500)$ and  $f_0(980)$ appear in relative $L=0$ meson-meson orbital angular momentum, and then $\pi \pi$  in the final state selects $I=0$; hence, the $\pi^0 \eta$ intermediate state does not contribute.

One final element of information is needed to complete the formula for $d\Gamma/dM_{inv}$, with $M_{inv}$ the $\pi^+ \pi^-$ or $\pi K$ invariant mass, which is the fact that we need an $L'=1$ orbital angular momentum for the $J/\psi$  in a $0^- \to 1^- 0^+$ transition to match angular momentum conservation. In \cite{weihong}  we assumed $V_P = A~p_{J/\psi} \cos \theta$, although the combination with spin produces a different angular dependence. In practice, the explicit form does not matter since it is the same for the different reactions, and we only care about the ratios between them. The only thing that matters is the presence of the factor $p_{J/\psi}$. Thus, we follow the formalism of \cite{weihong} and write
\begin{equation}\label{eq:dGamma}
  \frac{d \Gamma}{d M_{inv}}=\frac{1}{(2\pi)^3}\frac{1}{4M_{\bar B_j}^2}\frac{1}{3}p_{J/\psi}^2 p_{J/\psi} \tilde{p}_{\pi} {\overline{ \sum}} \sum \left| \tilde{t}_{\bar B^0_j \to J/\psi \pi^+ \pi^-} \right|^2,
\end{equation}
where the factor $1/3$ is coming from the integral of $\cos^2 \theta$ and $\tilde{t}_{\bar B^0_j \to J/\psi \pi^+ \pi^-}$ is $t_{\bar B^0_j \to J/\psi \pi^+ \pi^-}/(p_{J/\psi} \cos \theta)$, which depends on the $\pi^+ \pi^-$ invariant mass. In
Eq.~(\ref{eq:dGamma}), $p_{J/\psi}$ is the $J/\psi$ momentum in the global CM frame ($\bar B$ at rest), and $\tilde{p}_{\pi}$ is the pion momentum in the $\pi^+ \pi^-$ rest frame,
\begin{equation}\label{eq:pJpsi}
 p_{J/\psi}=\frac{\lambda^{1/2}(M_{\bar B}^2, M_{J/\psi}^2, M_{inv}^2)}{2M_{\bar B}},~~~~~~~\tilde{p}_{\pi}=\frac{\lambda^{1/2}(M_{inv}^2, m_{\pi}^2, m_{\pi}^2)}{2M_{inv}}.
\end{equation}
The formulas for the $\pi K$ invariant mass distribution are similar to Eqs. (\ref{eq:dGamma}) and (\ref{eq:pJpsi}), but with $M_{inv}$ being the $\pi K$ invariant mass and substituting one of the $m_{\pi}^2 $
in $~\tilde{p}_{\pi}$ by $m_K ^2 $.

\section{Formalism for vector meson production}

The diagrams of Fig. \ref{fig:fig1} without the hadronization can serve to study the production of vector mesons, which are largely $q \bar{q}$ states \cite{pelaez, acetirho, acetiks}. Since we were concerned up to now only about ratio of the scalars, the factor $V_{P}$ was taken arbitrary. In order to connect the scalar meson production with the vector production, we need a factor $V_{H}$ associated to the hadronization. Here, instead, the spin of the particles requires $L'=0,2$, and with no rule preventing $L'=0$, we assume that it is preferred; hence, the $p_{J/\psi} \cos\theta$ is not present now. Then we find immediately the amplitudes associated to Fig.  \ref{fig:fig1},
\begin{align}
&t_{\bar{B}^{0} \rightarrow J/\psi \rho^{0}}= -\frac{1}{\sqrt{2}}\tilde{V}^{\prime}_{P}~ V_{cd},
~&&t_{\bar{B}^{0} \rightarrow J/\psi \omega}= \frac{1}{\sqrt{2}}\tilde{V}'_{P} ~V_{cd},
~&&t_{\bar{B}^{0}_{s} \rightarrow J/\psi \phi}= \tilde{V}'_{P} ~V_{cs}, \nonumber \\
&t_{\bar{B}^{0} \rightarrow J/\psi \bar{K}^{*0}}= \tilde{V}'_{P} ~V_{cs},
~&&t_{\bar{B}^{0}_{s} \rightarrow J/\psi K^{*0}}= \tilde{V}'_{P} ~V_{cd},
\label{eq:tmatM}
\end{align}
where $(-\frac{1}{\sqrt{2}})$ is the $\rho^{0}$ component in $d \bar{d}$ and $(\frac{1}{\sqrt{2}})$ that of the $\omega$.  In order to determine $\tilde{V}'_{P}$ versus $\tilde{V}_{P}$ in the scalar production, we use the well-measured ratio\cite{LHCb:2012ae,pdg},

\begin{equation}
  \frac{\Gamma_{\bar{B}^{0}_{s}\rightarrow J/\psi f_{0} (980);f_{0} (980)\rightarrow \pi^{+} \pi^{-}} }{\Gamma_{\bar{B}^{0}_{s} \rightarrow J/\psi \phi}}=(13.9 \pm 0.9) \times 10^{-2}.
\label{eq:BR1}
\end{equation}

The width for $J/\psi V$ vector decay is now given by
\begin{equation}
\Gamma_{V_{i}}=\frac{1}{8 \pi}\frac{1}{m_{\bar B^{0}_i}^2} \left| t_{\bar B^0_i \to J/\psi V_{i}}\right|^2 p_{J/ \psi}.
  \label{eq:Gam1}
\end{equation}
Equations (\ref{eq:tmatM}) allow us to determine ratios of vector production with respect to the $\phi$,
\begin{eqnarray}
 && \frac{\Gamma_{\bar{B}^{0}\rightarrow J/\psi \rho^{0}}}{\Gamma_{\bar{B}^{0}_{s} \rightarrow J/\psi \phi}}= \frac{1}{2} \left| \dfrac{V_{cd}}{V_{cs}} \right|^2  \dfrac{m_{\bar{B}^{0}_{s}}^{2}}{m_{\bar{B}^{0}}^{2}} \dfrac{p_{\rho^{0}}}{p_{\phi}} =0.0263, \nonumber\\
 && \frac{\Gamma_{\bar{B}^{0}\rightarrow J/\psi \omega}}{\Gamma_{\bar{B}^{0}_{s} \rightarrow J/\psi \phi}}= \frac{1}{2} \left| \dfrac{V_{cd}}{V_{cs}} \right|^2  \dfrac{m_{\bar{B}^{0}_{s}}^{2}}{m_{\bar{B}^{0}}^{2}} \dfrac{p_{\omega}}{p_{\phi}} =0.0263,  \\
  && \frac{\Gamma_{\bar{B}^{0}\rightarrow J/\psi \bar{K}^{*0}}}{\Gamma_{\bar{B}^{0}_{s} \rightarrow J/\psi \phi}}=  \dfrac{m_{\bar{B}^{0}_{s}}^{2}}{m_{\bar{B}^{0}}^{2}} \dfrac{p_{\bar{K}^{*0}}}{p_{\phi}} =0.957, \nonumber\\
   &&\frac{\Gamma_{\bar{B}_{s}^{0}\rightarrow J/\psi K^{*0}}}{\Gamma_{\bar{B}^{0}_{s} \rightarrow J/\psi \phi}}=  \left| \dfrac{V_{cd}}{V_{cs}} \right|^2   \dfrac{p_{ K^{*0}}}{p_{\phi}} =0.0551.\nonumber
  \label{eq:ratio1}
  \end{eqnarray}

 By taking as input the branching ratio of $\bar{B}^{0}_{s} \rightarrow J/\psi \phi$,
  \begin{equation}
BR(\bar{B}^{0}_{s} \rightarrow J/\psi \phi)=(10.0^{+3.2}_{-1.8})\times 10^{-4},
  \label{eq:BR2}
\end{equation}
we obtain the other four branching ratios
\begin{eqnarray}
&&BR(\bar{B}^{0} \rightarrow J/\psi \rho^{0})=(2.63^{+0.84}_{-0.47})\times 10^{-5}, \nonumber\\
&&BR(\bar{B}^{0} \rightarrow J/\psi \omega)=(2.63^{+0.84}_{-0.47})\times 10^{-5}, \nonumber\\
&&BR(\bar{B}^{0} \rightarrow J/\psi \bar{K}^{*0})=(9.57^{+3.1}_{-1.7})\times 10^{-4}, \nonumber\\
&&BR(\bar{B}_{s}^{0} \rightarrow J/\psi K^{*0})=(5.51^{+1.7}_{-1.0})\times 10^{-5}.
  \label{eq:BR3}
\end{eqnarray}
The experimental numbers are \cite{pdg}
\begin{eqnarray}
&&BR(\bar{B}^{0} \rightarrow J/\psi \rho^{0})=(2.58 \pm 0.21)\times 10^{-5}, \nonumber\\
&&BR(\bar{B}^{0} \rightarrow J/\psi \omega)=(2.3 \pm 0.6)\times 10^{-5}, \nonumber\\
&&BR(\bar{B}^{0} \rightarrow J/\psi \bar{K}^{*0})=(1.34\pm 0.06)\times 10^{-3}, \nonumber\\
&&BR(\bar{B}_{s}^{0} \rightarrow J/\psi K^{*0})=(4.4\pm 0.9)\times 10^{-5}.
  \label{eq:BR4}
\end{eqnarray}
We can see that the agreement is good within errors, taking into account that the only theoretical errors in Eq. (\ref{eq:BR3}) are from the experimental branching ratio of Eq. (\ref{eq:BR2}). In the case of $BR(\bar{B}^{0} \rightarrow J/\psi \bar{K}^{*0})$, the agreement is borderline because of the small experimental errors. Admitting only $5 \% $ extra error from the theory, the agreement is quite good. Note also that the experimental $BR$ for $\bar{B}^{0} \rightarrow J/\psi \bar{K}^{*0}$ of the PDG has abnormally small errors; the most recent measurement from $BABAR$ gives $(1.33^{+0.22}_{-0.21}) \times 10^{-3}$ \cite{babarks}.

The rates discussed above have also been evaluated using perturbative QCD in the factorization approach in \cite{Liu:2013nea}, with good agreement with experiment.  Our approach has exploited flavor symmetries and the dominance of the weak decay mechanisms of Fig. 1 to calculate ratios of rates with good accuracy in a very easy way. Yet, our main purpose is to relate these rates with those for the production of scalar mesons to which we come below.

The next step is to compare the $\rho$ production with $\rho \rightarrow \pi^{+} \pi^{-}$ decay with $\bar{B}^{0} \rightarrow J/\psi f_{0}; f_{0}\rightarrow \pi^{+} \pi^{-} ( f_{0} \equiv f_{0}(500) ,f_{0}(980) $). In an experiment that looks for $\bar{B}^{0} \rightarrow J/\psi \pi^{+} \pi^{-}$, all these contributions will appear together, and only a partial wave analysis will disentangle the different contributions. This is done in \cite{Aaij:2013zpt, Aaij:2014siy}. There (see Fig. 13 of \cite{ Aaij:2014siy}) one observes a peak of the $\rho$ and an $f_{0}(500)$ distribution with a peak of the $\rho^{0}$ distribution about a factor $6$ larger than that of the $f_{0}(500)$. The $f_{0}(980)$ signal is very small and not shown in the figure.

In order to compare the theoretical results with these experimental distributions, we convert the rates obtained in Eqs. (\ref{eq:BR3}) into $\pi^{+} \pi^{-}$ distributions for the case of the $\bar{B}^{0} \rightarrow J/\psi \rho^{0}$ decay and $K^{-} \pi^{+}$ for the case of the $\bar{B}^{0} \rightarrow J/\psi \bar{K}^{*0}$ decay. For this purpose, we multiply the decay width of the $\bar{B}^{0}$ by the spectral function of
\begin{equation}
  \frac{d \Gamma_{\bar{B}^{0} \rightarrow J/\psi \rho^{0}}}{d M_{inv}(\pi^{+} \pi^{-})}=-\frac{1}{\pi}2 M_{\rho}~{\rm Im} \dfrac{1}{M_{inv}^{2}-M_{\rho}^{2}+i~M_{\rho} \Gamma_{\rho}(M_{inv}) } \Gamma_{\bar{B}^{0} \rightarrow J/\psi \rho^{0}} ,
 \label{eq:ratrho}
\end{equation}
where
\begin{eqnarray}
  &&\Gamma_{\rho}(M_{inv})= \Gamma_{\rho} \left (\dfrac{p_{\pi}^{\rm off}}{p_{\pi}^{\rm on}} \right )^{3}, \nonumber\\
  &&p_{\pi}^{\rm off}=\dfrac{\lambda^{1/2}(M_{inv}^{2},m_{\pi}^{2},m_{\pi}^{2})}{2M_{inv}} \theta (M_{inv}-2m_{\pi}), \nonumber\\
  &&p_{\pi}^{\rm on}=\dfrac{\lambda^{1/2}(M_{\rho}^{2},m_{\pi}^{2},m_{\pi}^{2})}{2M_{\rho}}.
 \label{eq:ratrhovwhere}
\end{eqnarray}
For the case of the $\bar{B}^{0} \rightarrow J/\psi \bar{K}^{*0} ~(\bar{K}^{*0}\rightarrow  \pi^{+} K^{-})$,
we have
\begin{equation}
  \frac{d \Gamma_{\bar{B}^{0} \rightarrow J/\psi \bar{K}^{*0};\bar{K}^{*0}\rightarrow  \pi^{+} K^{-}}}{d M_{inv}(\pi^{+} K^{-})}=-\frac{1}{\pi}\frac{2}{3}~{\rm Im} \dfrac{2 M_{K^{*}}}{M_{inv}^{2}-M_{K^{*}}^{2}+i~M_{K^{*}} \Gamma_{K^*}(M_{inv}) } \Gamma_{\bar{B}^{0} \rightarrow J/\psi \bar{K}^{*0}} ,
 \label{eq:KbarStr}
\end{equation}
with
\begin{eqnarray}
  &&\Gamma_{K^{*}}(M_{inv})= \Gamma_{K^{*}} \left (\dfrac{p^{\rm off}}{p^{\rm on}} \right )^{3},\nonumber\\
  &&p^{\rm off}=\dfrac{\lambda^{1/2}(M_{inv}^{2},m_{\pi}^{2},m_{K}^{2})}{2M_{inv}} \theta (M_{inv}-m_{\pi}-m_{K}),\nonumber\\
  &&p^{\rm on}=\dfrac{\lambda^{1/2}(M_{K^{*}}^{2},m_{\pi}^{2},m_{K}^{2})}{2M_{K^{*}}},
 \label{eq:KbarStrw}
\end{eqnarray}
and similarly for $\bar B^0_s \to J/\psi K^{*0}; K^{*0} \to \pi^- K^+$.

In Eqs. (\ref{eq:ratrho}) and (\ref{eq:KbarStr}) we have taken into account that $\rho^{0}$ decays only in $\pi^{+} \pi^{-}$, while $ \bar{K}^{*0} $ decays into $\pi^{+} K^{-}$, $\pi^{0} \bar{K}^{0}$ with weights $2/3$ and $1/3$, respectively.

\section{Results}
\begin{figure}[ht!]\centering
\includegraphics[height=7.0cm,keepaspectratio]{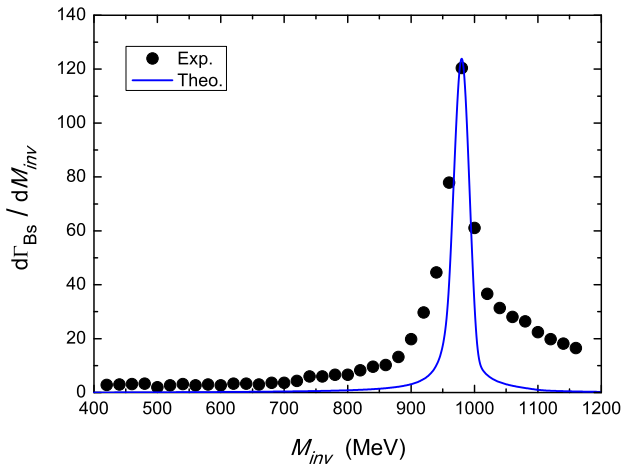}
\caption{$\pi^+ \pi^-$ invariant mass distribution for the $\bar B^0_s \to J/\psi \pi^+ \pi^-$ decay, with arbitrary normalization. Data from \cite{Aaij:2014emv}.}
\label{fig:dGammaBs}
\end{figure}

In Fig. \ref{fig:dGammaBs} we show our results for  $\bar B_{s}^0 \to J/\psi \pi^+ \pi^-$ decay. $\tilde{V}_{P}$ has been taken equal to 1 in this arbitrary normalization. The factor $\tilde{V}'_{P}$ of  Eqs. (\ref{eq:tmatM})
has then been adjusted to get the ratio of Eq. (\ref{eq:BR1}).
One can see a clear signal for $\bar B_{s}^0 \to J/\psi f_{0}(980), f_{0}(980) \rightarrow \pi^+ \pi^-$. It is also clear that there is no appreciable signal for $f_{0}(500)$ production as observed in the experiment \cite{Aaij:2014emv}. This is a clean case since the $q \bar{q}$ produced was $s \bar{s}$, which has $I=0$ and there is no $\rho^{0}$ production.

\begin{figure}[ht!]\centering
\includegraphics[height=7.5cm,keepaspectratio]{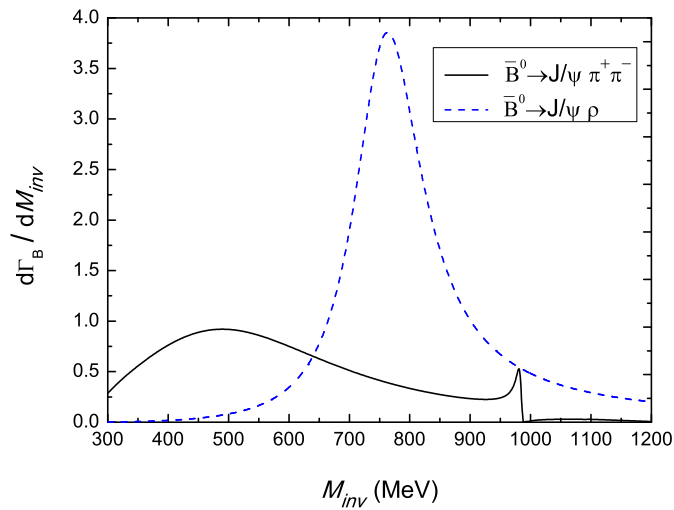}
\caption{$\pi^+ \pi^-$ invariant mass distributions for the $\bar B^0 \to J/\psi \pi^+ \pi^-$ ($S$ wave) (solid line) and $\bar B^0 \to J/\psi \rho$, $\rho \to \pi^+ \pi^- $ ($P$ wave) decays, with arbitrary normalization.}
\label{fig:dGammaB}
\end{figure}

In Fig. \ref{fig:dGammaB} we show our predictions for $f_{0}(500)$, $f_{0}(980)$, and $\rho^{0}$ production in  $\bar B^0 \to J/\psi \pi^+ \pi^-$, with the same normalization as in Fig.  \ref{fig:dGammaBs}.

The relative strengths and the shapes of the  $f_{0}(500)$ and $\rho$  distributions are remarkably similar to those found in the partial wave analysis of \cite{Aaij:2014siy}. However, our $f_{0}(500)$ has a somewhat different shape since in the analysis of \cite{Aaij:2014siy}, like in many experimental papers, a Breit-Wigner shape for the $f_{0}(500)$ is assumed, which is different to what the $\pi \pi$ scattering and the other production reactions demand \cite{zousigma,osetli}.

\begin{figure}[ht!]\centering
\includegraphics[height=7.5cm,keepaspectratio]{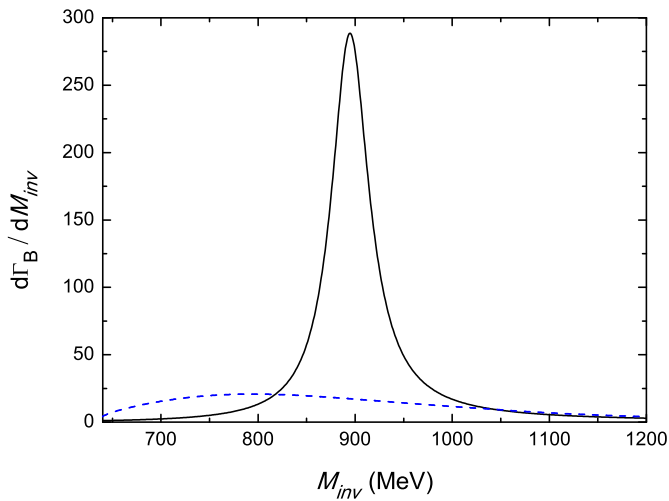}
\caption{ $\pi^+ K^-$ invariant mass distributions for the $\bar B^0 \to J/\psi \bar K^{*0}$, $\bar K^{*0} \to \pi^+ K^-$ (solid line) and $\bar B^0 \to J/\psi \bar \kappa$, $\bar \kappa \to \pi^+ K^-$ (dashed line), with arbitrary normalization.}
\label{fig:Kap1}
\end{figure}

\begin{figure}[ht!]\centering
\includegraphics[height=7.5cm,keepaspectratio]{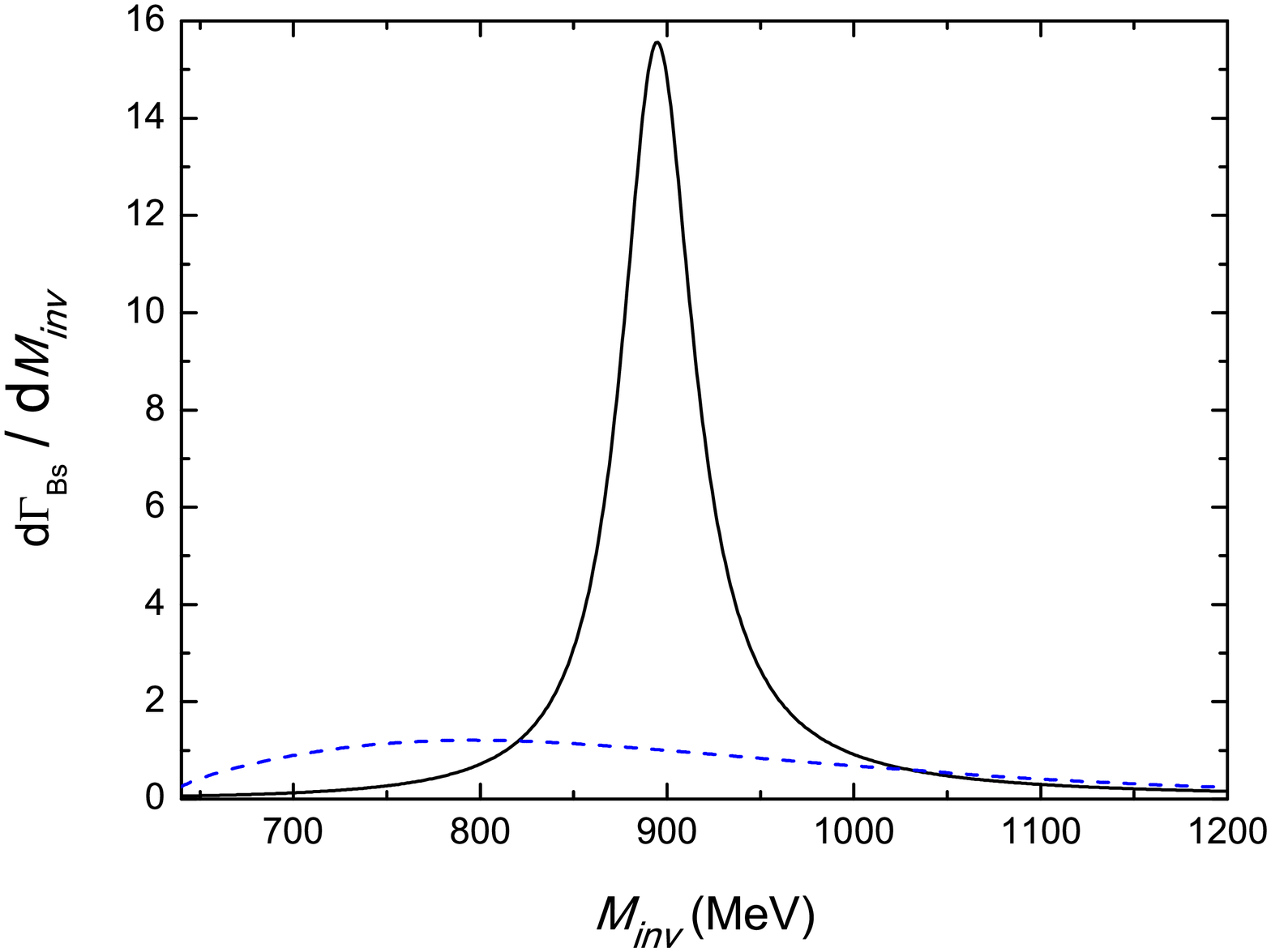}
\caption{ $\pi^- K^+$ invariant mass distributions for the $\bar B^0_s \to J/\psi K^{*0}$, $K^{*0} \to \pi^- K^+$ (solid line) and $\bar B^0_s \to J/\psi \kappa$, $\kappa \to \pi^- K^+$ (dashed line), with arbitrary normalization.}
\label{fig:Kap2}
\end{figure}

In Fig. \ref{fig:Kap1} we show the  results for  the Cabbibo allowed $\bar B^0 \to J/\psi \pi^+ K^-$, superposing  the contribution of the $\bar \kappa$ and $\bar K^{*0}$ contributions and in Fig. \ref{fig:Kap2}  the results for the Cabbibo suppressed $\bar B_{s}^0 \to J/\psi \pi^- K^+$, with the contributions of $\kappa$ and $K^{*0}$.

The narrowness of the  $K^{*}$ relative to the $\rho$, makes the wide signal of the scalar $\kappa$ to show clearly in regions where the $K^{*0}$ strength is already suppressed. While no explicit mention of the  $\kappa$ resonance is done in these $\bar B$ decays, in some analyses, a background is taken that resembles very much the  $\kappa$ contribution that we have in Fig. \ref{fig:Kap1} \cite{Abe:2002haa}. The  $\kappa(800)$ appears naturally in chiral unitary theory of $\pi K$ and coupled channel scattering as a broad resonance around $800$ MeV, similar to the $f_0(500)$ but with strangeness \cite{ramonet}. In $D$ decays, concretely in the $D^{+} \rightarrow K^- \pi^{+} \pi^{+}$ decay, it is studied with attention and the links to chiral dynamics are stressed \cite{Aitala:2002kr,Pennington:2007se}. With the tools of partial wave analysis developed in \cite{Aaij:2014siy}, it would be interesting to give attention to this $S$-wave resonance in future analysis.

\section{Discussion and conclusion}
  In this paper, we have addressed the problem of the $\bar B^0$ and $\bar B^0_s$ decays into $J/\psi$ $f_0(980)$,  $J/\psi$ $f_0(500)$, and $J/\psi$ $\kappa(800)$. In addition, we have also studied the decay of these $B$ states into  $J/\psi$ and a vector meson, $\rho$, $\omega$, $\phi$, $K^{*0}$, $\bar K^{*0}$. We have isolated the dominant mechanism for the weak decay of the $B$ meson, going to $J/\psi$ and a $q \bar q$ pair. This mechanism already allows us to relate the different vector decays,
$J/\psi$ and $\rho$, $\omega$, $\phi$, $K^{*0}$, $\bar K^{*0}$, with a good agreement with experiment for the four predictions that we can make. The production of the scalar mesons is more subtle since it requires the hadronization of the $q \bar q$ pair into a pair of mesons. We have implemented this step, and after that the pair of mesons are allowed to interact with their coupled channels, and this interaction generates the low-lying scalar resonances $f_0(500)$,  $f_0(980)$,and $\kappa(800)$. By using the experimental rate of $J/\psi$ $f_0(980)$ production from the   $\bar B^0_s$ decay versus the one to $J/\psi$ $\phi$, we can convert all the ratios of rates obtained into absolute numbers. We have compared the results with experiment and found good agreement with experiment for the different observables. In particular, the ratio of $\rho$ production to $f_0(500)$ production in $\bar B^0$ decay is in fair agreement with the results of a recent partial wave analysis of data.  We also have made predictions for  $J/\psi$ $\kappa$ production versus $J/\psi$ $K^{0*}$ and $J/\psi$ $\bar K^{0*}$ that can be tested in further partial wave analysis of these decays.

At this point, we would like to have a discussion concerning the nature of the scalar mesons, a topic of permanent debate, as briefly addressed in the Introduction. The light scalars $f_0(500)$, $f_0(980)$, $a_0(980)$, $\kappa (800)$ discussed here have been associated in the past to $q \bar q$ structures \cite{torn}, a $K \bar K$ molecule in the case of the  $f_0(980)$ \cite{isgur}, and a $(q)^2 (\bar q)^2$ tetraquark \cite{Jaffe:1976ih} among others (see reviews \cite{Klempt:2007cp,Crede:2008vw}). The advent of chiral unitary theory, the line followed in the present work, offered a new perspective where these light scalars were a consequence of chiral dynamics and unitarity in coupled channels: the chiral Lagrangian provides a potential between pseudoscalar mesons, which used in the context of the Bethe-Salpeter equations in coupled channels, leads to poles in the scattering matrix which correspond to the physical states. Since they are generated from the multiple scattering of the mesons via the  Bethe-Salpeter equation, they are called dynamically generated states and would correspond to meson-meson molecular states in the ordinary classification. The discussion on this issue keeps going. In \cite{polosa} these mesons are considered as $(q)^2 (\bar q)^2$ tetraquarks. In \cite{amir} they are considered as a mixture of $q \bar q$ and tetraquarks, and the $f_0(500)$ would have about 40\% probability of $q \bar q$ and 60\% of tetraquark. It is clear that the debate is not over. Yet, in view of this, let us go back to the chiral unitary approach and reinterpret the previous results. We can see a link of the quark pictures with the chiral unitary approach by recalling the interesting observation made in \cite{amir} that "the properties obtained within these quark models should be interpreted as being 'bare' properties, subject to a nontrivial renormalization due to effects which provide unitary corrections to the scattering amplitudes in which these particles appear as poles." These unitary corrections, incorporating meson-meson multiple scattering, are so huge that upon its implementation the width of the $f_0(500)$ appears of the order of 500 MeV, as seen in the work of \cite{amirscatt}. Certainly, this effect of the meson-meson scattering adds necessarily meson-meson components to the wave function of the $f_0(500)$, that in terms of quarks could be interpreted as having largely a four-quark component, or a meson-meson component if we use a different basis to write the wave function. This effect was already discussed in \cite{Tornqvist:1995ay,van Beveren:1986ea}, where starting from a seed of $q \bar q$, its coupling to meson-meson components was considered and unitarization was implemented, concluding that the meson cloud took over the original seed of $q \bar q$ in the light scalar mesons. In the chiral unitary approach, one already makes a starting point from this position, implicitly assuming that these states are meson-meson composite states, but recalling that whatever model one wishes to do for the scalars, the explicit consideration of the meson-meson multiple scattering is unavoidable if one wishes to have an accurate description of experimental data. This said, the approach is rather flexible in the sense that, even with its strong predictive power, there is one parameter that one has to fit to experiment (some scattering data usually), which is the regularization parameter in the meson-meson loop functions (a cutoff, or a subtraction constant if dimensional regularization is used). It is well known that, in coupled channels,
the effects of less important missing channels
can be largely accounted for by a suitable change of the cutoff or subtraction constant \cite{Oller:2000ma,weihong}.  In this sense, the approach is flexible enough to account for possible $q \bar q$ components in the wave function which are not explicitly considered in the different meson-meson components of the wave function. It is from this perspective that one should view the success of the chiral unitary approach describing different reactions where the light scalars are produced, as briefly discussed in the Introduction and in \cite{weihong}. Ultimately, the suitability of different pictures to describe experimental phenomena should be the guiding principle into the discussion of the nature of the scalars. The present work brings a new step in this direction. Steps in the same direction with different pictures would certainly be most welcome.

   In the discussion about the nature of hadrons, in which the vector mesons stand as largely $q \bar q$ states while the low-lying scalar mesons are rather dynamically generated states from the meson-meson interaction, we have shown that the $B$ decays investigated here greatly support this picture. We studied together the two decay modes into the $J/\psi$ scalar and $J/\psi$ vector from this perspective, and we obtained a remarkable agreement with experimental results which range in several orders of magnitude.

\section*{Acknowledgments}

We would like to thank Diego Milanes for useful discussions and motivation to do this work,  and S. Stone and L. Zhang for enlightening discussions concerning the experiments. This work is partly supported by the Spanish Ministerio de Economia
y Competitividad and European FEDER funds under the contract number
FIS2011-28853-C02-01 and FIS2011-28853-C02-02, and the Generalitat
Valenciana in the program Prometeo II, 2014/068. We acknowledge the
support of the European Community-Research Infrastructure
Integrating Activity Study of Strongly Interacting Matter (acronym
HadronPhysics3, Grant Agreement n. 283286) under the Seventh
Framework Programme of EU. This work is also partly supported by the
National Natural Science Foundation of China under Grant No. 11165005. This work is also partly supported by TUBITAK under the project No. 113F411.

\bibliographystyle{plain}

\begin{thebibliography}{999}

\bibitem{Aaij:2011fx}
  R.~Aaij {\it et al.}  [LHCb Collaboration],
  Phys.\ Lett.\ B {\bf 698}, 115 (2011)
  [arXiv:1102.0206 [hep-ex]].

\bibitem{Li:2011pg}
  J.~Li {\it et al.}  [Belle Collaboration],
  Phys.\ Rev.\ Lett.\  {\bf 106}, 121802 (2011)
  [arXiv:1102.2759 [hep-ex]].



\bibitem{Aaltonen:2011nk}
  T.~Aaltonen {\it et al.}  [CDF Collaboration],
  Phys.\ Rev.\ D {\bf 84}, 052012 (2011)
  [arXiv:1106.3682 [hep-ex]].

\bibitem{Abazov:2011hv}
  V.~M.~Abazov {\it et al.}  [D0 Collaboration],
  Phys.\ Rev.\ D {\bf 85}, 011103 (2012)
  [arXiv:1110.4272 [hep-ex]].


\bibitem{LHCb:2012ae}
  RAaij {\it et al.}  [LHCb Collaboration],
  Phys.\ Rev.\ D {\bf 86}, 052006 (2012)
  [arXiv:1204.5643 [hep-ex]].



\bibitem{Aaij:2014emv}
  R.~Aaij {\it et al.}  [LHCb Collaboration],
  Phys.\ Rev.\ D {\bf 89}, 092006 (2014)
  [arXiv:1402.6248 [hep-ex]].

\bibitem{Aaij:2013zpt}
  RAaij {\it et al.}  [LHCb Collaboration],
  Phys.\ Rev.\ D {\bf 87}, no. 5, 052001 (2013)
  [arXiv:1301.5347 [hep-ex]].

\bibitem{Colangelo:2010bg}
  P.~Colangelo, F.~De Fazio and W.~Wang,
  Phys.\ Rev.\ D {\bf 81}, 074001 (2010)
  [arXiv:1002.2880 [hep-ph]].



\bibitem{Stone:2013eaa}
  S.~Stone and L.~Zhang,
  Phys.\ Rev.\ Lett.\  {\bf 111}, no. 6, 062001 (2013)
  [arXiv:1305.6554 [hep-ex]].

\bibitem{weihong}
  W.~H.~Liang and E.~Oset,
  Phys.\ Lett.\ B {\bf 737}, 70 (2014)
  [arXiv:1406.7228 [hep-ph]].

\bibitem{npa}
  J.~A.~Oller and E.~Oset,
  Nucl.\ Phys.\ A {\bf 620}, 438 (1997)
  [Erratum-ibid.\ A {\bf 652}, 407 (1999)]
  [hep-ph/9702314].

\bibitem{ramonet}
  J.~A.~Oller, E.~Oset and J.~R.~Pelaez,
  Phys.\ Rev.\ D {\bf 59}, 074001 (1999)
  [Erratum-ibid.\ D {\bf 60}, 099906 (1999)]
  [Erratum-ibid.\ D {\bf 75}, 099903 (2007)]
  [hep-ph/9804209].

\bibitem{kaiser}
N.~Kaiser,
Eur.\ Phys.\ J.\ A {\bf 3}, 307 (1998).

\bibitem{markushin}
M.~P.~Locher, V.~E.~Markushin and H.~Q.~Zheng,
Eur.\ Phys.\ J.\ C {\bf 4}, 317 (1998)
[hep-ph/9705230].

\bibitem{juanito}
J.~Nieves and E.~Ruiz Arriola,
Nucl.\ Phys.\ A {\bf 679}, 57 (2000)
[hep-ph/9907469].


\bibitem{rios}
J.~R.~Pelaez and G.~Rios,
Phys.\ Rev.\ Lett.\ {\bf 97}, 242002 (2006)
[hep-ph/0610397].

\bibitem{Kaiser:1995eg}
  N.~Kaiser, P.~B.~Siegel and W.~Weise,
  Nucl.\ Phys.\ A {\bf 594}, 325 (1995)
  [nucl-th/9505043].

\bibitem{Oller:2000ma}
  J.~A.~Oller, E.~Oset and A.~Ramos,
  Prog.\ Part.\ Nucl.\ Phys.\  {\bf 45}, 157 (2000)
  [hep-ph/0002193].

\bibitem{Gasser:1983yg}
  J.~Gasser and H.~Leutwyler,
  Annals Phys.\  {\bf 158}, 142 (1984).

\bibitem{Bernard:1995dp}
  V.~Bernard, N.~Kaiser and U.~-G.~Meissner,
  Int.\ J.\ Mod.\ Phys.\ E {\bf 4}, 193 (1995)
  [hep-ph/9501384].


\bibitem{Meissner:2000bc}
  U.~-G.~Meissner and J.~A.~Oller,
  Nucl.\ Phys.\ A {\bf 679}, 671 (2001)
  [hep-ph/0005253].

\bibitem{chiangpalo}
  L.~Roca, J.~E.~Palomar, E.~Oset and H.~C.~Chiang,
  Nucl.\ Phys.\ A {\bf 744}, 127 (2004)
  [hep-ph/0405228].

\bibitem{Lahde:2006wr}
  T.~A.~Lahde and U.~-G.~Meissner,
  Phys.\ Rev.\ D {\bf 74}, 034021 (2006)
  [hep-ph/0606133].

\bibitem{Hanhart:2007bd}
  C.~Hanhart, B.~Kubis and J.~R.~Pelaez,
  Phys.\ Rev.\ D {\bf 76}, 074028 (2007)
  [arXiv:0707.0262 [hep-ph]].

\bibitem{Roca:2012cv}
  L.~Roca,
  Phys.\ Rev.\ D {\bf 88}, 014045 (2013)
  [arXiv:1210.4742 [hep-ph], arXiv:1210.4742 [hep-ph]].


\bibitem{robert}
   A.~Furman, R.~Kaminski, L.~Lesniak and B.~Loiseau,
   Phys.\ Lett.\ B {\bf 622}, 207 (2005)
   [hep-ph/0504116].


\bibitem{bruno}
   B.~El-Bennich, A.~Furman, R.~Kaminski, L.~Lesniak and B.~Loiseau,
   Phys.\ Rev.\ D {\bf 74}, 114009 (2006)
   [hep-ph/0608205].

\bibitem{Aaij:2012nh}
  R.~Aaij {\it et al.}  [LHCb Collaboration],
  Phys.\ Rev.\ D {\bf 86}, 071102 (2012)
  [arXiv:1208.0738 [hep-ex]].



\bibitem{pdg}
  J.~Beringer {\it et al.}  [Particle Data Group Collaboration],
  Phys.\ Rev.\ D {\bf 86}, 010001 (2012) and 2013 partial update for the 2014 edition (URL: http://pdg.lbl.gov)


\bibitem{Aubert:2007qea}
  B.~Aubert {\it et al.}  [BaBar Collaboration],
  Phys.\ Rev.\ D {\bf 76}, 092004 (2007)
  [arXiv:0707.1648 [hep-ex]].

\bibitem{Aubert:2004rz}
  B.~Aubert {\it et al.}  [BaBar Collaboration],
  Phys.\ Rev.\ Lett.\  {\bf 94}, 141801 (2005)
  [hep-ex/0412062].

\bibitem{Aaij:2014siy}
  R.~Aaij {\it et al.}  [LHCb Collaboration],
  Phys.\ Rev.\ D {\bf 90}, 012003 (2014)
  [arXiv:1404.5673 [hep-ex]].


\bibitem{alberzou}
  A.~Martinez Torres, L.~S.~Geng, L.~R.~Dai, B.~X.~Sun, E.~Oset and B.~S.~Zou,
  Phys.\ Lett.\ B {\bf 680}, 310 (2009)
  [arXiv:0906.2963 [nucl-th]].

\bibitem{danydan}
  D.~Gamermann, E.~Oset, D.~Strottman and M.~J.~Vicente Vacas,
  Phys.\ Rev.\ D {\bf 76}, 074016 (2007)
  [hep-ph/0612179].

\bibitem{pelaez}
  J.~R.~Pelaez,
  Phys.\ Rev.\ Lett.\  {\bf 92}, 102001 (2004)
  [hep-ph/0309292].


\bibitem{acetirho}
  F.~Aceti and E.~Oset,
  Phys.\ Rev.\ D {\bf 86}, 014012 (2012)
  [arXiv:1202.4607 [hep-ph]].

\bibitem{acetiks}
  C.~W.~Xiao, F.~Aceti and M.~Bayar,
  Eur.\ Phys.\ J.\ A {\bf 49}, 22 (2013)
  [arXiv:1210.7176 [hep-ph]].


\bibitem{babarks}
  B.~Aubert {\it et al.}  [BaBar Collaboration],
  Phys.\ Rev.\ D {\bf 76}, 092004 (2007)
  [arXiv:0707.1648 [hep-ex]].


\bibitem{Liu:2013nea}
  X.~Liu, W.~Wang and Y.~Xie,
  Phys.\ Rev.\ D {\bf 89}, 094010 (2014)
  [arXiv:1309.0313 [hep-ph]].


\bibitem{zousigma}
  M.~Ablikim {\it et al.}  [BES Collaboration],
  Phys.\ Lett.\ B {\bf 598}, 149 (2004)
  [hep-ex/0406038].


\bibitem{osetli}
  C.~-b.~Li, E.~Oset and M.~J.~Vicente Vacas,
  Phys.\ Rev.\ C {\bf 69}, 015201 (2004)
  [nucl-th/0305041].


\bibitem{Abe:2002haa}
  K.~Abe {\it et al.}  [Belle Collaboration],
  Phys.\ Lett.\ B {\bf 538}, 11 (2002)
  [hep-ex/0205021].


\bibitem{Aitala:2002kr}
  E.~M.~Aitala {\it et al.}  [E791 Collaboration],
  Phys.\ Rev.\ Lett.\  {\bf 89}, 121801 (2002)
  [hep-ex/0204018].

\bibitem{Pennington:2007se}
  J.~M.~Link {\it et al.}  [FOCUS Collaboration],
  Phys.\ Lett.\ B {\bf 653}, 1 (2007)
  [arXiv:0705.2248 [hep-ex]].


\bibitem{torn}
  N.~A.~Tornqvist,
  Z.\ Phys.\ C {\bf 68}, 647 (1995)
  [hep-ph/9504372].

\bibitem{isgur}
  J.~D.~Weinstein and N.~Isgur,
  Phys.\ Rev.\ Lett.\  {\bf 48}, 659 (1982).

\bibitem{Jaffe:1976ih}
  R.~L.~Jaffe,
  Phys.\ Rev.\ D {\bf 15}, 281 (1977).

\bibitem{Klempt:2007cp}
  E.~Klempt and A.~Zaitsev,
  Phys.\ Rept.\  {\bf 454}, 1 (2007)
  [arXiv:0708.4016 [hep-ph]].

\bibitem{Crede:2008vw}
  V.~Crede and C.~A.~Meyer,
  Prog.\ Part.\ Nucl.\ Phys.\  {\bf 63}, 74 (2009)
  [arXiv:0812.0600 [hep-ex]].

\bibitem{polosa}
  L.~Maiani, F.~Piccinini, A.~D.~Polosa and V.~Riquer,
  Phys.\ Rev.\ Lett.\  {\bf 93}, 212002 (2004)
  [hep-ph/0407017].

\bibitem{amir}
  A.~H.~Fariborz, R.~Jora and J.~Schechter,
  Phys.\ Rev.\ D {\bf 79}, 074014 (2009)
  [arXiv:0902.2825 [hep-ph]].



\bibitem{amirscatt}
  A.~H.~Fariborz, N.~W.~Park, J.~Schechter and M.~Naeem Shahid,
  Phys.\ Rev.\ D {\bf 80}, 113001 (2009)
  [arXiv:0907.0482 [hep-ph]].

\bibitem{Tornqvist:1995ay}
  N.~A.~Tornqvist and M.~Roos,
  Phys.\ Rev.\ Lett.\  {\bf 76} (1996) 1575
  [arXiv:hep-ph/9511210].

\bibitem{van Beveren:1986ea}
  E.~van Beveren, T.~A.~Rijken, K.~Metzger, C.~Dullemond, G.~Rupp and J.~E.~Ribeiro,
  Z.\ Phys.\  C {\bf 30} (1986) 615
  [arXiv:0710.4067 [hep-ph]].










\end{thebibliography}

\end{document}